\documentclass[aps, pra, reprint, twocolumn, 10pt, superscriptaddress]{revtex4-1}

\usepackage{amsmath, amssymb}
\usepackage{mathtools}
\usepackage{braket}
\usepackage{siunitx}
\usepackage{textcomp}
\usepackage{xcolor} 

\usepackage{math}
\usepackage{color}
\usepackage{hyperref}
\usepackage{braket}
\usepackage{siunitx}

\begin{document}

\title{Excitation of the \texorpdfstring{$^{229}$Th}{229Th}  nucleus via a two-photon electronic transition}

\author{Robert A. M\"uller}
\email{robert.mueller@ptb.de}
\affiliation{Physikalisch-Technische Bundesanstalt, D-38116 Braunschweig, Germany}
\affiliation{Technische Universit\"at Braunschweig, D-38106 Braunschweig, Germany}

\author{Andrey V. Volotka}
\affiliation{Helmholtz Insitute Jena, D-07743 Jena, Germany}
\affiliation{GSI Helmholtzzentrum f\"ur Schwerionenforschung, D-64291 Darmstadt, Germany}

\author{Andrey Surzhykov}
\affiliation{Physikalisch-Technische Bundesanstalt, D-38116 Braunschweig, Germany}
\affiliation{Technische Universit\"at Braunschweig, D-38106 Braunschweig, Germany}

\begin{abstract}
We investigate the process of nuclear excitation via a two-photon electron transition (NETP) for the case of the doubly charged thorium. The theory of the NETP process has been devised originally for heavy helium like ions. In this work we study this process in the nuclear clock isotope $^{229}$Th in the $2+$ charge state. For this purpose we employ a combination of configuration interaction and many-body perturbation theory to calculate the probability of NETP in resonance approximation. The experimental scenario we propose for the excitation of the low lying isomeric state in $^{229}$Th is a circular process starting with a two-step pumping stage followed by NETP. The ideal intermediate steps in this process depend on the supposed energy $\hbar\omega_N$ of the nuclear isomeric state. For each of these energies the best initial state for NETP is calculated. Special focus is put on the most recent experimental results for $\hbar\omega_N$.
\end{abstract}

\maketitle

\section{Introduction}
\begin{figure*}[htb]
\includegraphics[width=0.48\textwidth]{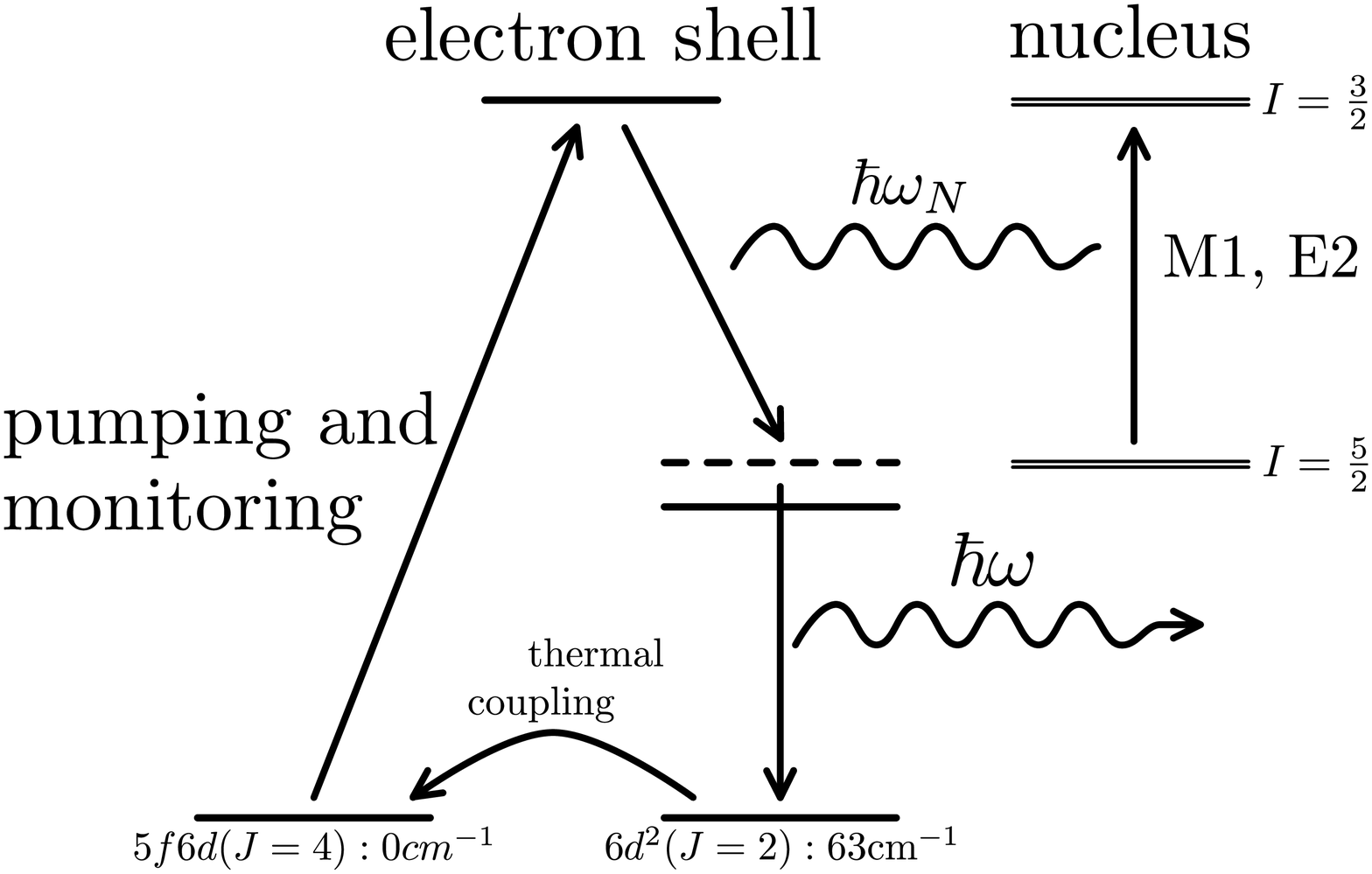}
\hfill
\includegraphics[width=0.48\textwidth]{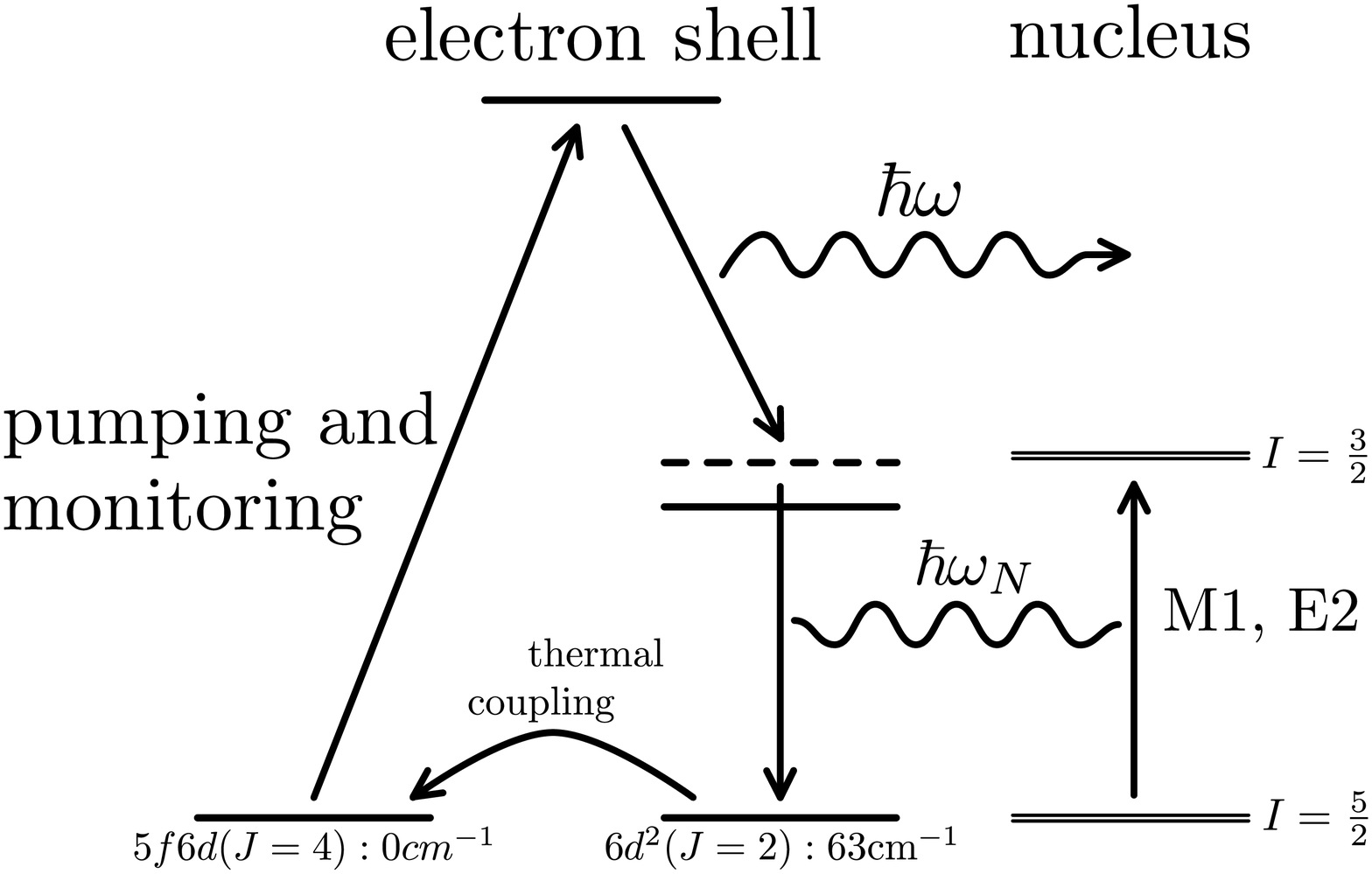}
\caption{Sketch of the NETP process together with the initial pumping of the excited state via two-step laser excitation. There are two equivalent scenarios to be considered; either the first (left picture) or the second photon (right picture) excites the nucleus. The excitation is assumed to start from the $6d^2(J=2):\SI{63}{cm^{-1}}$ state, which is assumed to always be thermally populated in the sample. \label{fig:process sketch}}
\end{figure*}
Atomic clocks are amongst the most precise measurement instruments available to date \cite{bloom_optical_2014, huntemann_single-ion_2016}. Accurate time measurements, clock comparisons offer the opportunity to investigate fundamental physics and possible physics beyond the standard model \cite{kostelecky_constraints_1999, rosenband_frequency_2008, safronova_highly_2014}. Fifteen years ago it has been proposed by Peik and coworkers to build a clock based on a nuclear transition \cite{peik_nuclear_2003}. The most suitable of such transitions is found in the thorium isotope with mass number $A=229$ between the nuclear ground and the first excited isomeric state, nowadays sometimes referred to as \emph{nuclear clock isomer}. Therefore intense research, theoretically and experimentally, has been performed on $^{229}$Th and especially the nucleus in its first excited state, the isomer $^{229m}$Th \cite{wense_direct_2017, tkalya_decay_2018, seiferle_lifetime_2017, safranova_nuclear_2018}. Recently, for example, the nuclear moments of $^{229m}$Th have been determined \cite{thielking_laser_2018, muller_hyperfine_2018}, which may give insight into the energy of the nuclear isomeric state \cite{beloy_hyperfine_2014}. Moreover the emission of internal conversion electrons from the $^{229m}$Th$\rightarrow^{229}$Th transition has been observed \cite{von_der_wense_direct_2016}. However a controlled excitation of the nuclear isomer has not been achieved yet \cite{stellmer_attempt_2018}.

A large number of different processes have been proposed to produce the $^{229m}$Th nuclear isomer ranging from direct laser excitation to the interaction with hot plasmas \cite{karpeshin_electron_2002, peik_nuclear_2003, palffy_theory_2006, gunst_dominant_2014, andreev_nuclear_2019, von_der_wense_laser_2017}. Out of these the excitation of nuclei by the energy excess from electronic processes appears to be very efficient and is much stronger than e.g. direct laser excitation \cite{gunst_dominant_2014, gunst_direct_2015, muller_hyperfine_2018}. However all such electronic bridge processes come with a major challenge: For the process to be sufficiently strong, the electronic transition needs to be in proximity to the transition between the nuclear ground and the low lying isomeric state of $^{229}$Th. In the classic electron bridge process the energy difference between the electronic and nuclear transition is accounted for by the absorption of a photon. Instead of absorbing a photon with an enery that needs to be precisely tuned we consider a two-photon decay in the electron shell \cite{volotka_nuclear_2016}. In such a transition one, virtual, photon excites the nucleus while the other is emitted as a real photon. The energy share between both photons is continuous and, thus, there is no scanning necessary to excite the nucleus. This so called nuclear excitation by a two-photon electron transition (NETP) has been introduced for heavy highly charged ions, to access nuclear excited states in the keV regime \cite{volotka_nuclear_2016}.

In this work we want to investigate NETP in $^{229m}$Th. In contrast to other nuclear levels, the $^{229m}$Th isomeric state is found only about $\SI{8}{eV}$ above the $^{229}$Th ground state. Therefore the electronic transition needs to be in the same energy range. Consequently lower charge states, especially $^{229}$Th$^{2+}$, are promising candidates to observe NETP in thorium.

In contrast to the scenario discussed in Ref. \cite{volotka_nuclear_2016} for helium like ions, Th$^{2+}$ has many real intermediate resonances between the upper and the final state of the NETP process, provided by the rich level structure of the thorium ion. Ideally such a resonance is close to the nuclear excitation energy, thus enhancing the probability of the NETP process. The location and number of the resonances, however, strongly depends on the initially pumped upper state. Therefore the upper state which offers the highest probability for NETP depends on the energy of the nuclear isomeric state. In this paper we therefore provide detailed calculations for NETP in $^{229}$Th$^{2+}$ and give clear recommendations for the levels to excite, depending on the energy range in which the isomer is searched.

Hartree atomic units ($\hbar=m_e=e=1$) are used throughout this paper unless stated otherwise.

\section{Scenario}
A sketch of the scheme we propose for the excitation of the low lying isomeric state in $^{229}$Th can be seen in Fig. \ref{fig:process sketch}. First, starting from the $5f6d(J=4):\SI{0}{cm^{-1}}$ ground state, the electron shell of the thorium ion is excited to an upper state with odd parity. From this upper state the NETP process occurs, where the nuclear excitation energy either corresponds to the energy splitting between the upper and the intermediate (left panel) or the intermediate and the lower state. The NETP decay is either of E1+M1 or E1+E2 type, therefore the final state of the process is of even parity. In this work we consider the $6d^2(J=2):\SI{63}{cm^{-1}}$ state as the final state, which is almost degenerate with the $5f6d(J=4):\SI{0}{cm^{-1}}$ ground state. In previous experiments on Th$^+$ and Th$^{2+}$ with a buffer gas quenched sample no significant population of dark states has been observed. Thus, we can safely assume that the $6d^2(J=2):\SI{63}{cm^{-1}}$ state decays quickly to the ground state due to collisional coupling \cite{knoop_collisional_1998, herrera-sancho_two-photon_2012, thielking_laser_2018, muller_hyperfine_2018, thielking_2019}.

\section{Theory}
\subsection{NETP Transition Amplitudes and Rates}
In the previous section we have described the process we propose for the excitation of $^{229}$Th. Now we will derive the probability of NETP in doubly charged thorium below. To simplify our considerations we will assume that the pumping of the upper state (cf. \ref{fig:process sketch}) is very efficient so that it is always populated. Therefore the probability of the process is given by the last two deexcitation steps which resemble the NETP process as discussed in Ref. \cite{volotka_nuclear_2016}.

In this work we will identify each many-electron state by its total angular momentum $J$, the projection $\mu$ of $J$ onto the quantization ($z$-) axis and a set of additional quantum numbers summarized by $\gamma$. The nuclear states are labelled by the nuclear spin $I$ and its projection $M$. The theoretical description of NETP consists of two interfering channels. As seen also in Fig. \ref{fig:process sketch} either the first or the second photon can excite the nucleus. Consequently the NETP matrix element $M_{fi}$ consists of two terms: 
\begin{widetext}
\begin{equation}
\begin{aligned}
M_{fi}=&\sum_{\gamma_n J_n \mu_n}\left(\frac{\braket{\gamma_f J_f \mu_f, I_e M_e|\vec{\alpha}\cdot\vec{u}_\lambda\e^{i\vk\cdot\vr}|\gamma_n J_n \mu_n, I_e M_e}\braket{\gamma_n J_n \mu_n,I_e M_e|H_{int}|\gamma_i J_i \mu_i,I_g M_g}}{\epsilon_i-\omega_N-\epsilon_n-\im\frac{\Gamma_n}{2}}\right.\\
&+\left.\frac{\braket{\gamma_f J_f \mu_f, I_e M_e|H_{int}|\gamma_n J_n \mu_n, I_g M_g}\braket{\gamma_n J_n \mu_n,I_g M_g|\vec{\alpha}\cdot\vec{u}_\lambda\e^{i\vk\cdot\vr}|\gamma_i J_i \mu_i,I_g M_g}}{\epsilon_f+\omega_N-\epsilon_n-\im\frac{\Gamma_n}{2}}\right),
\end{aligned}
\label{eq:matrix element}
\end{equation}
\end{widetext}
where $\omega_N$ is the frequency of the nuclear transition and $\vec{\alpha}$ denotes the vector of Dirac matrices. State energies and widths are denoted by $\epsilon$ and $\Gamma$, respectively, while the subscripts $i$, $n$ and $f$ specify the initial, intermediate and final states.  Generally the intermediate state can be virtual and, thus, we have to sum over the entire spectrum $\ket{\gamma_n J_n \mu_n}$, where we assume that the continuous spectrum can be neglected. Note that in Eq. \eqref{eq:matrix element} we have omitted the width of the nuclear excited state, since it is much narrower than the electronic states.

Both terms in Eq. \eqref{eq:matrix element} each split into two matrix elements of the operators $H_{int}$ and $\vec{\alpha}\cdot\vec{u}_\lambda\e^{i\vk\cdot\vr}$. The latter is the usual interaction of the electron shell with a plane-wave photon with momenteum $\vk$ polarized along $\vec{u}_\lambda$, where $\lambda$ is the helicity. The interaction Hamiltonian $H_{int}$ mediates the interaction between the electron shell and the nucleus, thus acting on both electronic and nuclear degrees of freedom.

To obtain the probability of the NETP process we can use Fermi's golden rule:
\begin{widetext}
\begin{equation}
\mathcal{W}_{fi}=\frac{\alpha^3\omega}{2\pi}\sum_{\mu_i\mu_f M_g M_e \lambda}[J_i,I_g]^{-1}\int|M_{fi}|^2\dd\Omega_k,
\label{eq:probability general}
\end{equation}
\end{widetext}
where $[k]=2k+1$, $\alpha$ is the fine structure constant, $\dd\Omega_k$ the differential emission angle of the real photon and $\omega$ the frequency of the real, emitted, photon. In Eq. \eqref{eq:probability general} we average over $M_g$ and $\mu_i$, assuming that the initial electronic and nuclear states are unpolarized. Moreover neither $\mu_f$ and $M_e$ nor the emission direction of the real photon is observed, thus we sum over the magnetic quantum numbers of the final states and integrate over $\Omega_k$.

To express Eq. \eqref{eq:probability general} in a more convenient way the photon emission operator $\vec{\alpha}\cdot\vec{u}_\lambda\e^{i\vk\cdot\vr}$ is readily expanded into electric ($p=1$) and magnetic ($p=0$) multipoles $L$ with magnetic quantum number $M$\cite{eisenberg_nuclear_1976, rose_elementary_1995}:
\begin{equation}
\vec{u}_\lambda\e^{i\vk\cdot\vr}=\sqrt{2\pi}\sum_{LMp}\im^L[L]^{\frac{1}{2}}(\im\lambda)^pD^L_{M\lambda}(\phi_k,\theta_k,0)\vec{a}^{(p)}_{LM},\\
\end{equation}
where $D_{M\lambda}^L(\phi_k, \theta_k, 0)$ is the Wigner-D matrix and $\vec{a}^{(p)}_{LM}$ are irreducible tensors of rank $L$ resembling the multipole fields.

Similar to the photon interaction operator, the electron-nucleus interaction $\hat{H}_{int}$ can be expanded into multipoles \cite{porsev_excitation_2010, porsev_electronic_2010}:
\begin{equation}
\hat{H}_{int}=\sum_{qr}\hat{T}_{qr}\hat{M}^*_{qr},
\end{equation}
where it is important to note that for each multipole the operator $\hat{H}_{int}$ splits into the hyperfine interaction operators $\hat{T}_{qr}$ acting only on electronic degrees of freedom and $\hat{M}_{qr}$ interacting with the nuclear part of the wave function. That way we can find the NETP probability for each multipolarity $q$ of the nuclear transition and electronic transitions $L$ and $p$.
\begin{equation}
\begin{aligned}
\mathcal{W}^{(Lpq)}_{fi}=&\frac{8\pi\alpha^3\omega}{[J_i,I_g]}\sum_{Lpq}\frac{|\braket{I_e||\hat{M}_{q}||I_g}|^2}{[q]}\\
&\times\left(G^{(Lpq)}_1+G^{(Lpq)}_2+G^{(Lpq)}_{12}\right),
\end{aligned}
\label{eq:probability multipoles}
\end{equation}
where the total probability of the process would be the sum over all possible $L$, $p$ and $q$ and
\begin{widetext}
\begin{subequations}
\begin{equation}
G^{(Lpq)}_1=\sum_{J_n}\frac{1}{[J_n]}\left|\sum_{\gamma_n} \frac{\braket{\gamma_f J_f||\vec{\alpha}\cdot\vec{a}^{(p)}_{L}||\gamma_n J_n}\braket{\gamma_n J_n||\hat{T}_{q}||\gamma_i J_i}}{\epsilon_i-\omega_N-\epsilon_n-\im\frac{\Gamma_n}{2}}\right|^2,
\end{equation}
\begin{equation}
G^{(Lpq)}_2=\sum_{J_n}\frac{1}{[J_n]}\left|\sum_{\gamma_n} \frac{\braket{\gamma_f J_f||\hat{T}_{q}||\gamma_n J_n}\braket{\gamma_n J_n||\vec{\alpha}\cdot\vec{a}^{(p)}_{L}||\gamma_i J_i}}{\epsilon_f+\omega_N-\epsilon_n-\im\frac{\Gamma_n}{2}}\right|^2,\\
\end{equation}
\begin{equation}
\begin{aligned}
G^{(Lpq)}_{12}=&2(-1)^{q+L}\sum_{\substack{\gamma_n \gamma_n'\\J_n J_n'}}(-1)^{J_n+J_n'}
\left\lbrace
\begin{matrix}
J_i & q & J_n\\
J_f & L & J_n'
\end{matrix}
\right\rbrace\mathrm{Re}\left(\frac{\braket{\gamma_f J_f||\vec{\alpha}\cdot\vec{a}^{(p)}_{L}||\gamma_n J_n}\braket{\gamma_n J_n||\hat{T}_{q}||\gamma_i J_i}}{\epsilon_i-\omega_N-\epsilon_n-\im\frac{\Gamma_n}{2}}\right.\\
&\times\left.\frac{\braket{\gamma_f J_f||\hat{T}_{q}||\gamma_n' J_n'}^*\braket{\gamma_n' J_n'||\vec{\alpha}\cdot\vec{a}^{(p)}_{L}||\gamma_i J_i}^*}{\epsilon_f+\omega_N-\epsilon_n'+\im\frac{\Gamma_n'}{2}}\right).
\end{aligned}
\end{equation}
\end{subequations}
\end{widetext}
The equations above show that the NETP probability for each multipole \eqref{eq:probability multipoles} splits into three parts proportional to the amplitudes $G^{(Lpq)}_i$. The first two amplitudes $G^{(Lpq)}_1$ and $G^{(Lpq)}_2$ correspond here to the cases illustrated in Fig. \ref{fig:process sketch}, where the real photon is emitted either due to the transition between the initial and the intermediate or the intermediate and the final state. The last amplitude $G^{(Lpq)}_{12}$ covers the interference between these two coherent processes.
\subsection{Resonance Approximation}
For our specific case the probability \eqref{eq:probability multipoles} can be further simplified. In contrast to the very simple electronic structure of helium-like systems, for which NETP has been first discussed \citep{volotka_nuclear_2016}, Th$^{2+}$ has a rich and dense level structure. Therefore it is safe to assume that only the closest resonance will contribute to the NETP probability. This allows for the application of the so-called resonance approximation. In this approximation all interference terms vanish, thus, $G^{(Lpq)}_{12}$ can be neglected and the terms $G^{(Lpq)}_{1}$ and $G^{(Lpq)}_{2}$ become:
\begin{subequations}
\begin{equation}
\begin{aligned}
G^{(Lpq)}_1\approx&\sum_{\gamma_n J_n}\frac{1}{[J_n]}\\
&\times\left|\frac{\braket{\gamma_f J_f||\vec{\alpha}\cdot\vec{a}^{(p)}_{L}||\gamma_n J_n}\braket{\gamma_n J_n||\hat{T}_{q}||\gamma_i J_i}}{\epsilon_i-\omega_N-\epsilon_n-\im\frac{\Gamma_n}{2}}\right|^2,
\end{aligned}
\end{equation}
\begin{equation}
\begin{aligned}
G^{(Lpq)}_2\approx&\sum_{\gamma_n J_n}\frac{1}{[J_n]}\frac{\Gamma_i+\Gamma_n}{\Gamma_n}\\
&\times\left|\frac{\braket{\gamma_f J_f||\hat{T}_{q}||\gamma_n J_n}\braket{\gamma_n J_n||\vec{\alpha}\cdot\vec{a}^{(p)}_{L}||\gamma_i J_i}}{\epsilon_f+\omega_N-\epsilon_n-\im\frac{\Gamma_i + \Gamma_n}{2}}\right|^2,
\end{aligned}
\end{equation}
\label{eq:G approx}
\end{subequations}
where we incorporated the width of the initial state in resonance approximation following Ref. \cite{shabaev_parity-nonconservation_2010}.

Now, the remaining task to calculate the NETP probability \eqref{eq:probability multipoles} in resonance approximation is the evaluation of the reduced nuclear and electronic matrix elements. The nuclear transition amplitudes $\braket{I_e||\hat{M}_{q}||I_g}$ are known from elaborate nuclear calculations, e.g. by Minkov and P\'alffy \cite{minkov_reduced_2017}, where previous estimates by Tkalya \emph{et al.} \cite{tkalya_radiative_2015} have been refined.
\subsection{Enhancement Factor \texorpdfstring{$\beta$}{beta}}
Due to the complexity of nuclear calculations, the nuclear amplitudes provided e.g. in Ref. \cite{minkov_reduced_2017} are a major source of uncertainty in our calculations of the NETP probability \eqref{eq:probability multipoles}. To circumvent these uncertainties one can define the \emph{enhancement factor} $\beta$ (cf. \cite{porsev_electronic_2010, porsev_excitation_2010}), which is independent on the nuclear transition probability:
\begin{equation}
\beta^{(Lpq)}=\frac{\mathcal{W}^{(Lpq)}_{fi}}{\Gamma_q},
\label{eq:beta}
\end{equation}
where the nuclear decay width $\Gamma_q$ is defined by:
\begin{equation}
\begin{aligned}
\Gamma_q&=\frac{8\pi(q+1)}{q((2q+1)!!)^2}\frac{(\alpha\omega_N)^{2q+1}}{[I_g]}\left|\braket{I_e||\hat{M}_{q}||I_g}\right|^2.
\end{aligned}
\end{equation}
The enhancement factor \eqref{eq:beta} is defined in analogy to Refs. \cite{porsev_electronic_2010, porsev_excitation_2010} and given here mainly to make a connection to these works and to test our theory with respect to effects coming from the electronic structure of Th$^{2+}$.

Specifically for the case of $^{229}$Th, the leading multipoles of the nuclear transition are $M1$ and $E2$, so $q$ is either $1$ or $2$. From now on we will assume that all radiative electronic transitions are of $E1$ type, so that $L=1$ and $p=1$. Therefore, in resonance approximation, the enhancement factors of interest are
\begin{subequations}
\begin{equation}
\beta^{(111)}=\frac{3\omega}{2\omega_N^3}\frac{1}{[J_i]}\left(G^{(111)}_1+G^{(111)}_2\right),
\end{equation}
\begin{equation}
\beta^{(112)}=\frac{30\omega}{\alpha^2\omega_N^5}\frac{1}{[J_i]}\left(G^{(112)}_1+G^{(112)}_2\right).
\end{equation}
\end{subequations}
\section{Numerical details}
Up to now we have shown how the NETP process may be discussed by taking the nuclear transition amplitude from the literature or by investigating the enhancement factor instead. Now we will briefly sketch the evaluation of the electronic matrix elements. To calculate these, we apply a combination of configuration interaction (CI) and many-body perturbation theory (MBPT), that has been described in detail in Refs. \cite{dzuba_combination_1996, dzuba_v_2005, dzuba_core-valence_2007}. In particular we used the package assembled by Kozlov \emph{et al.} \cite{kozlov_ci-mbpt:_2015}. The CI configuration state functions have been set up using Dirac-Hartree-Fock wave functions for the core orbitals and the $5f$, $6d$, $7s$ and $7p$ valence orbitals. For the higher lying orbitals we use a $b$-splines and orbitals constructed using the method described e.g. in Ref. \cite{kozlov_manifestation_1996}. The CI basis is constructed by virtual excitations from the $[Rn]+6d^2$ and $[Rn]+5f6d$ configuration.

The CI+MBPT method is a powerful method to calculate reliable transition matrix elements. Level energies, however, especially for complicated systems like Th$^{2+}$ are determined more accurately in experiments. Because the exact position of the resonances is important to determine the NETP probability accurately, we take the experimental values \cite{kramida_nist_1999} for all level energies instead of the theoretical ones. We will discuss the importance of this step in the section below.
\section{Results and Discussion}
\begin{figure*}[htb]
\includegraphics[width=\textwidth]{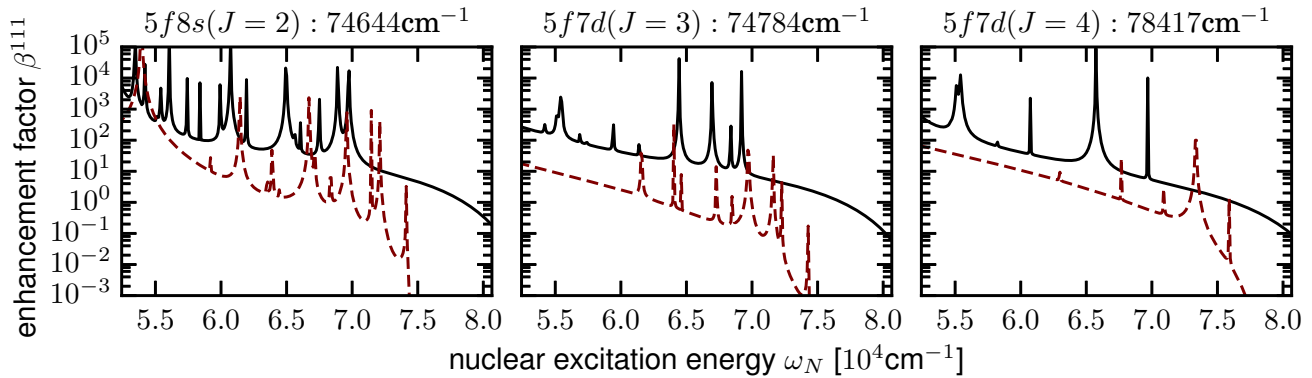}
\caption{Comparison of the enhancement factor $\beta$ before (black solid line) and after (red dashed line) the resonance energies have been shifted to the experimental values \cite{kramida_nist_1999}. Exemplarily shown here are the transitions from the $5f8s(J=2):\SI{74644}{cm^{-1}}$ (left panel), $5f8s(J=3):\SI{74784}{cm^{-1}}$ (center panel) and $5f7d(J=4):\SI{78417}{cm^{-1}}$ (right panel) states to the $6d^2(J=2):\SI{63}{cm^{-1}}$ state.\label{fig:beta}}
\end{figure*}
Before we discuss the probability of the NETP process in Th$^{2+}$, we will have a brief look on the enhancement factor $\beta^{(111)}$ [cf. Eq. \eqref{eq:beta}]. In particular we want to investigate how the replacement of the calculated level energies by the experimental ones influences the results. Therefore we performed calculations for $\beta^{(111)}$ as a function of the nuclear excitation energy $\omega_N$ using both. The results of these calculations are shown in Fig. \ref{fig:beta}, the theoretical (black solid line) and the experimental (red dashed line) level energies. The first feature we notice in Fig. \ref{fig:beta} is the different number of resonance peaks for different $J_i$ of the upper (initial) state. This can be explained by the sheer number of available decay paths to the $6d^2(J=2):\SI{63}{cm^{-1}}$ state from each of these upper states. While for $J_i=4$ and a $E1$ radiative transition, the intermediate state must have $J_n=3$, for $J_i=2$ there are three possible $J_n$ and, therefore, more intermediate resonances available. But there are two more important things to notice. Foremost we see that the high energy cutoff of $\beta$ is reduced for the case of the experimental level energies. Therefore we note, that it is very important to take the energy splitting between the initial and final electronic state accurately into account. Moreover we see that the replacement of the energies of the intermediate states to their experimental values does not change the qualitative behaviour of $\beta^{(111)}$ and, thus, can be safely done to achieve accurate results.
\begin{figure}[htb]
\includegraphics[width=0.48\textwidth]{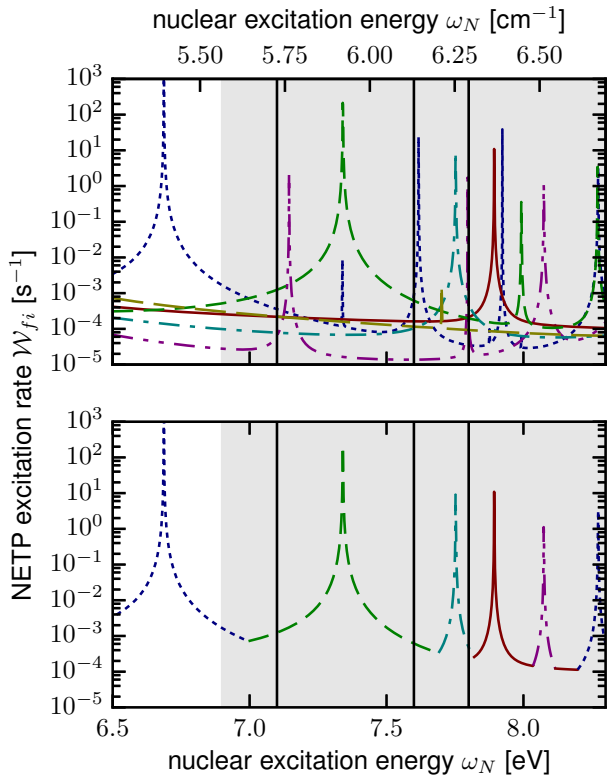}
\caption{Probability of NETP for different upper states (top panel) and the envelope of these probabilities (bottom panel), where contributions narrower than $\SI{0.1}{eV}$ are neglected. The colors are distributed as follows: blue dotted line: $5f8s(J=2):\SI{74644}{cm^{-1}}$, green dashed line: $5f7d(J=2):\SI{79916}{cm^{-1}}$, turqoise dash-dotted line: $5f7d(J=2):\SI{83237}{cm^{-1}}$, red solid line: $5f7d(J=3):\SI{84374}{cm^{-1}}$ and purple dash-double dotted line: $5f7d(J=2):\SI{78333}{cm^{-1}}$. The black vertical lines show the supposed energies of the low lying isomeric state according to Refs. \cite{beck_energy_2007, tkalya_radiative_2015, borisyuk_excitation_2018} with the corresponding uncertainty interval shown by the grey-shaded area.\label{fig:W_fi_zoom}}
\end{figure}
\begin{figure}[htb]
\includegraphics[width=0.48\textwidth]{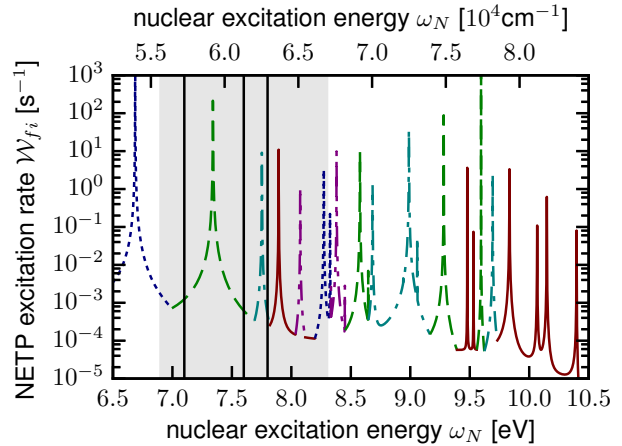}
\caption{Same as Fig. \ref{fig:W_fi_zoom} (bottom panel) but for the energy range between $\SI{6.5}{eV}$ and $\SI{10.5}{eV}$.\label{fig:W_fi}}
\end{figure}

The primary aim of this paper is to provide information about the most promising excitation paths to observe the NETP process in $^{229}$Th$^{2+}$. Therefore we assume according to available experimental setups that the exciting lasers are tunable between $\SI{3.45}{eV}$ and $\SI{5.25}{eV}$ \cite{meier_2018} (cf. Fig. \ref{fig:process sketch}). With such lasers $19$ possible upper states $\ket{\gamma_i J_i}$ can be pumped. This number reduces to $16$, if we fix the final state to be the level $6d^2(J=2):\SI{63}{cm^{-1}}$, in order to be able to cycle through the process multiple times. For each these $16$ possible upper states we calculated the NETP probability \eqref{eq:probability multipoles} summing over $1\leq q \leq 2$ in order to account for both the $M1$ and $E2$ nuclear transition channels. This step is necessary because it has been shown recently that both, the $M1$ and the $E2$ channel, may contribute equally to the NETP probability \cite{bilous_electric_2018}. Similar to Fig. \ref{fig:beta}, we display the NETP probability $\mathcal{W}_{fi}=\sum_{q=1}^2\mathcal{W}^{(E1,q)}_{fi}$ as a function of the nuclear excitation energy $\omega_N$. This data, however, is not very conclusive. Thus it needed to be processed, which is illustrated in Fig. \ref{fig:W_fi_zoom}. In the upper panel of this figure we display the NETP probability for four upper states as a function of the nuclear excitation energy $\omega_N$. To get our final result we take the envelope of this family of curves as shown in the bottom panel of Fig. \ref{fig:W_fi_zoom}. Moreover we omit resonance peaks narrower than $\SI{0.1}{eV}$ for it would make the figure impractical to use, especially at higher $\omega_N$, where the resonances get more dense. Note also that we do not show the NETP probability for those of the $16$ possible upper states that do not contribute to the envelope. The vertical lines in Fig. \ref{fig:W_fi_zoom} denote the most recent values for the energy of the nuclear isomeric state, ranging from $\SI{7.1}{eV}$ to $\SI{7.6}{eV}$ and $\SI{7.8}{eV}$ \cite{beck_energy_2007, tkalya_radiative_2015, borisyuk_excitation_2018}. The grey shaded area denotes the combined uncertainties of all three measurements and, thus, a recommended initial search area for the nuclear isomer.

With the preparation of the data explained above we are now able to generate the main result of this work. In Fig. \ref{fig:W_fi} we show, which is the ideal upper state to observe the NETP process in $^{229}$Th$^{2+}$ as a function of $\omega_N$. It can be seen that also for the entire energy range between $\SI{6.5}{eV}$ and $\SI{10.5}{eV}$ only $5$ of the possible $16$ upper states need to be considered for a possible experiment. Again the vertical lines and the grey area in Fig. \ref{fig:W_fi} mark the recommended initial search area for the nuclear isomeric state.

Let us finally discuss how the excitation of the nucleus could be monitored in the experiment we propose. Recently the hyperfine structure of the electronic levels in Th$^{2+}$ has proven to be a good indicator of whether the nucleus is in its ground or first excited state \cite{thielking_laser_2018}. This would be as well possible in the scenario proposed in the present work by either applying an additional laser or observing the fluorescence from one of the pumping stages. Another common option would be to observe the time delayed photoemission from the nuclear decay. This, however, would not be recommended for the scenario proposed here, because we could cycle through the process, no matter if the upper state decayed via NETP or the more likely two-photon cascade. This allows for a good statistics and does not require a shot-by-shot analysis of the data with accurate timing.

\section{Concluding remarks}

The NETP process has been shown to be a promising candidate to investigate the nuclear structure of highly charged ions \cite{volotka_nuclear_2016}. In the present work this process is discussed for many electron systems within the resonance approximation. To excite the $^{229}$Th nucleus, we propose a combination of a two-step pumping of an upper state from which the NETP process occurs. To overcome the difficulty of a small branching ratio between NETP and a generic radiative two-step decay of the upper state, the proposed process can be cycled independent on the way the ion decays.

With several promising experiments at the horizon that aim for a precise determination of $\hbar\omega_N$ \cite{von_der_wense_laser_2017, seiferle_towards_2018}, the scenario described in this work aims for a controlled excitation of the $^{229}$Th. Therefore a challenge that comes with many proposed electronic bridge processes for the excitation of the $^{229}$Th nucleus, the requirement of a continuous scanning with a tunable laser, does not apply in the scenario described in this paper. In the proposed experiment the lasers are adjusted only once to ensure the most efficient pumping of the upper state. For a first test of our theory we recommend to pump the $5f7d(J=2):\SI{83237}{cm^{-1}}$ and the $5f7d(J=3):\SI{84374}{cm^{-1}}$ states, which both have resonances close to the currently assumed value $\SI{7.8}{eV}$ of the nuclear excitation energy.

\begin{acknowledgments}
RAM acknowledges support form the RS-APS and many useful discussions with David-Marcel Meier and Johannes Thielking.
\end{acknowledgments}
\clearpage
\bibliography{quellen.bib}
\end{document}